\documentclass[aoas,MSNbibl,nameyear,seceqn,dvips]{arximspdf}
\usepackage{dcolumn}
\usepackage{graphicx}

\doi{10.1214/13-AOAS650} 
\volume{7}
\issue{3}
\pubyear{2013}
\firstpage{1334}
\lastpage{1361}

\makeatletter
\newcolumntype{d}[1]{D{.}{.}{#1}}
\newtheorem{prop}{Proposition}
\def\LOG10{{\log_{10}}}
\def\LOG2{{\log_{2}}}
\def\M{\mathbf{M}}

\def\bTheta{\bolds{\Theta}}
\def\bSigma{\bolds{\Sigma}}

\def\tN{{\mathrm{N}}}

\def\E{{\mathrm{E}}}
\def\tT{{\mathrm{T}}}
\def\P{\mathbf{P}}
\def\Q{\mathbf{Q}}
\def\w{\mathbf{w}}

\def\bgamma{\bolds{\gamma}}
\def\bxi{\bolds{\xi}}
\def\Z{\mathbf{Z}}
\def\z{\mathbf{z}}
\def\one{\mathbf{1}}

\def\bigo{{\mathcal{O}}}
\makeatother

\begin{document}
\begin{frontmatter}

\title{Bayesian clustering of replicated time-course gene expression data with weak signals}
\runtitle{Bayesian clustering}

\begin{aug}
\author[a]{\fnms{Audrey Qiuyan} \snm{Fu}\corref{}\thanksref{t1}\ead[label=e1]{audreyqyfu@uchicago.edu}},
\author[b]{\fnms{Steven}~\snm{Russell}\thanksref{t1}\ead[label=e2]{s.russell@gen.cam.ac.uk}},\break
\author[c]{\fnms{Sarah~J.}~\snm{Bray}\thanksref{t1}\ead[label=e3]{sjb32@cam.ac.uk}}
\and
\author[d]{\fnms{Simon} \snm{Tavar\'{e}}\thanksref{t1,t2}\ead[label=e4]{st321@cam.ac.uk}}
\thankstext{t1}{Supported in part by BBSRC Grant BBF00897X.}
\thankstext{t2}{Supported in part by NIH Grant P50 HG002790.}
\runauthor{Fu, Russell, Bray and Tavar\'{e}}
\affiliation{University of Cambridge, University of Cambridge,\break
University of Cambridge, and University of Cambridge and\break University of
Southern California}
\address[a]{A. Q. Fu\\
Department of Physiology, Development \\
\quad and Neuroscience\\
University of Cambridge\\
Downing Street\\
Cambridge CB2 3DY\\
United Kingdom\\
and\\
Cambridge Systems Biology Centre\\
Tennis Court Road\\
Cambridge CB2 1QR\\
United Kingdom\\
Current address:\\
Department of Human Genetics\\
University of Chicago\\
920 E 58th Street\\
Chicago, Illinois 60637\\
USA\\
\printead{e1}\vspace*{24pt}}

\address[b]{S. Russell\\
Department of Genetics\\
University of Cambridge\\
Downing Street\\
Cambridge CB2 3EH\\
United Kingdom\\
and\\
Cambridge Systems Biology Centre\\
Tennis Court Road\\
Cambridge CB2 1QR\\
United Kingdom\\
\printead{e2}}

\address[c]{S. J. Bray\\
Department of Physiology, Development \\
\quad and Neuroscience\\
University of Cambridge\\
Downing Street\\
Cambridge CB2 3DY\\
United Kingdom\\
\printead{e3}}

\address[d]{S. Tavar\'{e}\\
Department of Applied Mathematic\\
\quad and Theoretical Physics\\
University of Cambridge\\
Centre for Mathematical Sciences\\
Wilberforce Road\\
Cambridge CB3 0WA\\
United Kingdom\\
and\\
Program in Molecular \\
\quad and Computational Biology\\
University of Southern California\\
1050 Childs Way, RRI 201B\\
Los Angeles, California 90089-2910\\
USA\\
\printead{e4}}
\end{aug}

\received{\smonth{9} \syear{2012}}
\revised{\smonth{3} \syear{2013}}

%
\begin{abstract}
To identify novel dynamic patterns of gene expression, we develop a
statistical method to cluster noisy measurements
of gene expression collected from multiple replicates at
multiple time points, with an unknown number of clusters. We propose a
random-effects mixture model
coupled with a Dirichlet-process prior for clustering. The mixture
model formulation allows for probabilistic cluster assignments. The
random-effects formulation allows for attributing the total variability
in the data to the sources that are consistent with the experimental design,
particularly when the noise level is high and the temporal dependence
is not strong.
The Dirichlet-process prior induces a prior distribution on partitions
and helps to estimate the number of clusters (or mixture components)
from the data. We further tackle two challenges associated with
Dirichlet-process prior-based methods. One is efficient sampling. We develop
a novel Metropolis--Hastings Markov Chain Monte Carlo (MCMC) procedure
to sample the partitions. The other is efficient use of
the MCMC samples in forming clusters. We propose a two-step procedure
for posterior inference, which involves resampling and relabeling,
to estimate the posterior allocation probability matrix. This matrix
can be directly used in cluster assignments, while describing the uncertainty
in clustering. We demonstrate the effectiveness of our model and
sampling procedure through simulated data. Applying our method to
a real data set collected from \textit{Drosophila} adult muscle cells
after five-minute Notch activation, we identify 14 clusters of
different transcriptional
responses among 163 differentially expressed genes, which provides
novel insights into underlying transcriptional mechanisms in
the Notch signaling pathway. The algorithm developed here is
implemented in the R package DIRECT, available on CRAN.
\end{abstract}

%

\end{frontmatter}

\section{Introduction}
\label{sec:intro}
We are interested in the dynamics of the transcriptional response to
activation of the Notch signaling pathway~[\citet{Housden.etal.2012}].
During transcription, RNA molecules are produced using the DNA sequence
of the genes as templates, leading to the notion of these genes
being ``expressed.'' Some of the RNA molecules, mRNA specifically, are
subsequently translated into proteins, which directly regulate
all kinds of biological processes.
The highly conserved Notch signaling pathway mediates communication
between neighbouring cells. Located on the cell surface, the Notch
protein receives signals from adjacent cells and releases an
intracellular protein fragment that, along with other proteins, elicits
changes in
gene expression in the receiving cell.
Critical to the normal development of many organisms, the Notch
signaling pathway is under active
and extensive investigation [see \citet{Bray2006} for a review].

Using \textit{Drosophila}
as a model system, we aim to characterise patterns of the
transcriptional responses of the whole genome following a pulse of
Notch activation~[\citet{Housden.etal.2012}].
Previous studies have examined the changes in transcription at a single
time-point following Notch pathway activation~[\citet
{Jennings.etal.1994, Krejci.etal.2009}].
However, it is unclear
whether the regulated genes can have different temporal profiles, and
whether there are particular patterns of up-regulation (increased expression)
or down-regulation (decreased expression) amongst the genes whose
expression changes. To generate the data we
analyse here, Notch signaling was initiated in \textit{Drosophila} adult
muscle cells and stimulated for a short pulse of 5 minutes, and mRNA
levels were measured in these treated cells relative to untreated
cells, using microarrays for 4 biological replicates
at 18 unevenly-spaced time points during the 150 minutes after
activation [\citet{Housden2011,Housden.etal.2012}; also see
Section~\ref{sec:cases} for details on the experiment and
preprocessing of the data]. We aim to address the following questions
for the 163 differentially expressed genes: (i) how many different
expression patterns are there and
what are these patterns? and (ii) which genes exhibit what expression
pattern? These questions naturally call for a clustering
approach to analyse these data.

\begin{figure}

\includegraphics{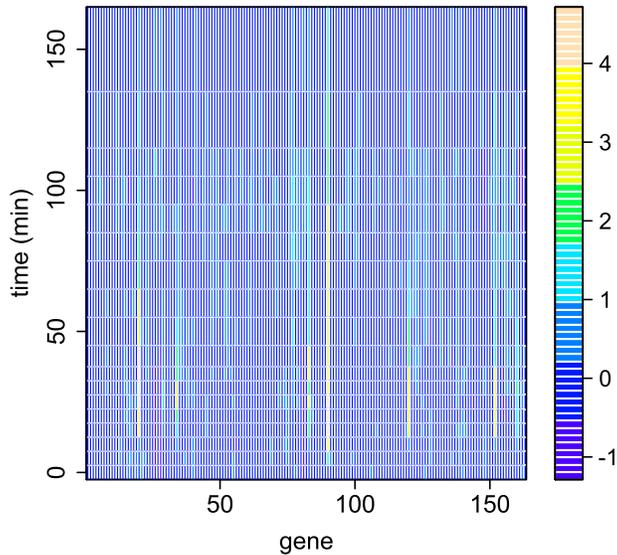}

\caption{Mean profiles of 163 significantly expressed genes [false
discovery rate 10$\%$ by EDGE; Storey et~al. (\citeyear{Storey.etal.2005})] over the time
course of 18 time points.
Each value is the mean (taken over the four replicates) of $\LOG2$ fold
change in treated cells relative to untreated cells.}
\label{fig:de163.mean}
\end{figure}

However, there are several challenges associated with this data set.
First, these data are different from the conventional time series
data. Time series often refer to the measurements of a single subject
over time. In the
microarray experiment, a~biological replicate refers to a \emph
{population} of cells, and the expression levels
at any time point are measured for a \emph{distinct sample} of cells
from the starting population. Although the cells
from the same biological replicate are typically assumed to be
homogeneous, the heterogeneity among cells is nonnegligible and contributes
to the noise in the data~[\citet{Spudich1976, McAdams1997,
Elowitz.etal.2002}]. Second, since only a short pulse of Notch activation
was applied, the level of (relative) expression, measured as $\LOG
2$-transformed fold change, in our data is often not much different
from 0
(Figure~\ref{fig:de163.mean}). Specifically, the mean expression
level across time points and across replicates is only 0.1 with a
standard deviation of 0.5, leading to a signal-to-noise ratio of only
$\sim$0.2.
Meanwhile, the median of the lag 1 autocorrelation across replicates is
only 0.4 (interquartile range: 0.2--0.6), indicating that the temporal
dependence is weak. Third, existing clustering software programs such
as MCLUST~[\citet{FraleyRaftery2002, FraleyRaftery2006}] and
SplineCluster~[\citet{Heard.etal.2006}] give vastly different results
(see Section~\ref{sec:cases} for detail).

These scientific questions and the challenges in the data thus
motivated the clustering method we develop here.
Our clustering method consists mainly of a random-effects mixture model
coupled with a Dirichlet-process prior. We propose
the random-effects model to tackle the high level of noise in the data
that arises from several sources. Under the
random-effects model, we make use of the full data, rather than
reducing the data to the means across replicates,
which may not be accurate with this level of noise. Under this model,
we also do not make many assumptions about
the underlying biological process, which is still largely unknown.
Novel patterns detected this way are unlikely to be the result of
potentially inappropriate assumptions. The use of a Dirichlet-process
prior enables us to estimate the number of
clusters directly from the data. Below we review existing relevant
work, which laid the foundation for our method.

Most clustering methods that are shown to be effective on time-course
data are model-based, with the distribution
following a mixture of multivariate Gaussian components~[\citet
{FraleyRaftery2002, MedvedovicSivaganesan2002, Medvedovic.etal.2004,
Celeux.etal.2005, BealKrish2006, FraleyRaftery2006, Heard.etal.2006,
Ma.etal.2006, Qin2006, ZhouWakefield2006, LauGreen2007,
Booth.etal.2008, Rasmussen.etal.2009, McNicholasMurphy2010, Green2010,
Cooke.etal.2011}].
Different methods take different approaches to modeling the mean
vectors and covariance structures.
Several methods attempt to account specifically for the temporal
dependence by modeling
the (prior) mean vector in terms of spline functions~[\citet
{Heard.etal.2006, Ma.etal.2006}] or as a random walk~[\citet{ZhouWakefield2006}].
As for the covariance structure, some methods [\citet
{MedvedovicSivaganesan2002, Medvedovic.etal.2004, Heard.etal.2006,
Qin2006, LauGreen2007, Green2010}]
assume independence across items and across time points a priori.
Both \citet{FraleyRaftery2006} and \citet{McNicholasMurphy2010} take a
matrix decomposition approach
and consider various models for the covariance matrix by constraining
no or some decomposed terms to be identical across
clusters. However, whereas \citet{FraleyRaftery2006} apply eigenvalue
decomposition, which is applicable also to data types
other than time-course data, \citet{McNicholasMurphy2010} use a modified
Cholesky decomposition, which has connections
with autoregressive models and is thus specifically designed for
time-course data. Another common approach to modeling
the covariance structure is
random-effects models, which account for variability arising from
different sources~[\citet{Celeux.etal.2005, Ma.etal.2006, Booth.etal.2008}].
We take this approach in our clustering method. Indeed, with a
random-effects mixture model, we demonstrate
that specific modeling of the temporal structure may not be essential
for clustering replicated time-course data.

Estimating the number of clusters, or mixture components, under a
model-based framework, has been a difficult problem.
Several approaches exist, largely falling into two categories:
optimization for a single ``best'' partition and a fully Bayesian
approach that weights the partitions by their probabilities given the
data. In the optimization category, the penalised likelihood approach,
using criteria such as the Akaike Information Criterion (AIC), Bayesian
Information Criterion (BIC) and so on, has been used
by~\citet{FraleyRaftery2002}, \citet{Celeux.etal.2005}, \citet
{Schliep.etal.2005}, \citet{Ma.etal.2006} and \citet
{McNicholasMurphy2010}. \citet{Heard.etal.2006} in their
program SplineCluster and \citet{Booth.etal.2008} maximise the posterior
probability of partitions given the data.
Arguing that the maximal posterior probability of partitions may be
difficult to compute reliably and may not be representative,
\citet{LauGreen2007} suggest maximizing posterior loss, an approach
followed in \citet{Green2010}. However, the main
drawback with the optimization approach
is that competing partitions with similar (penalised) likelihoods are
simply ignored. Methods based on optimization may
also suffer from numeric instability, as our experience with
MCLUST~[\citet{FraleyRaftery2002}] suggests (explained in Section~\ref{sec:cases}).
When clustering is used as an exploratory
data analysis tool to understand the heterogeneity in the data, it is
often desirable and realistic to explore more than
one partition and to understand how and why the data support multiple
competing partitions.
We therefore find the fully Bayesian approach more appealing with this
rationale. In this category,
\citet{ZhouWakefield2006} implemented the Birth-Death Markov Chain Monte
Carlo (BDMCMC) scheme
initially developed by \citet{Stephens2000a}, which
designs a birth-death process to generate new components and eliminate existing
ones. \citet{MedvedovicSivaganesan2002}, \citet{Medvedovic.etal.2004},
\citet{BealKrish2006}, \citet{Qin2006}, \citet{Booth.etal.2008} and \citet
{Rasmussen.etal.2009}
developed Markov Chain Monte Carlo (MCMC) schemes under a Dirichlet-process
prior. The Dirichlet-process prior, a popular tool in nonparametric
Bayesian statistics, can induce sparse partitions
among items~[\citet{Ferguson1973, Antoniak1974}] and has been widely
used in analyses such as nonparametric density
estimation~[\citet{EscobarWest1995, Fox2009}]. Here, we take the fully
Bayesian approach and use a Dirichlet-process prior
to induce a prior distribution on partitions, which helps us to explore
different numbers of clusters and to sample from partitions
under each number. The clustering obtained from
the Bayesian approach is essentially an average of all possible
solutions weighted by their posterior probabilities.

However, two major challenges remain in the clustering methods under
the Dirichlet-process priors.
One is efficient sampling. Many MCMC methods have been developed under
Dirichlet-process priors
for conjugate priors of the parameters [such as those reviewed in \citet
{Neal2000}], restricting
the choices of priors. Alternative sampling methods have been developed,
such as Gibbs \mbox{samplers} designed for nonconjugate priors~[\citet
{MacEachernMuller1998}],
several Metropolis--Hastings (MH) samplers under the Chinese-restaurant
representation~[\citet{Neal2000}],
split-merge sampling~[\citet{JainNeal2004, JainNeal2007}], another
two-stage MH procedure under an implicit
Dirichlet-process prior~[\citet{Booth.etal.2008}],
retrospective sampling [\citet{PapaRoberts2008}] and slice
sampling~[\citet{Walker2007, Kalli.etal.2011}],
both of which
are developed under the stick-breaking process representation.\vadjust{\goodbreak} Several
of these and related methods are reviewed recently in~\citet{GriffinHolmes2010}.
Here, we develop a novel MH sampler under the Chinese-restaurant
representation. Our MH sampler does not introduce additional
(auxiliary or latent) variables or tuning parameters.
It also does not require separate split and merge steps, but rather
allows for dimension changes in a single step. In addition, it is based on
standard MH calculations and is therefore straightforward to understand
and easy to implement.

The other major challenge is posterior inference. Existing approaches
[\citet{MedvedovicSivaganesan2002, Medvedovic.etal.2004, BealKrish2006,
Rasmussen.etal.2009, Dhavala.etal.2011}] attempt to
make use of the posterior
``similarity'' matrix, whose entries are the posterior probability of
allocating two items to the same cluster, by applying linkage-based
clustering algorithms to this matrix. Focusing on
this matrix in effect converts the original clustering problem into
another one, while discarding other valuable information in the MCMC
samples. We propose a two-step posterior inference procedure that
involves resampling and relabeling to estimate the posterior allocation
probability matrix,
which may be used more directly in forming clusters and other inference.

In this paper, we present our method DIRECT, the Dirichlet
process-based random-effects model as a clustering tool. We
describe the random-effects mixture model in Section~\ref{sec:model} and
the Bayesian inference in Section~\ref{sec:inference}, which includes a
novel MH MCMC algorithm for sampling partitions under the
Dirichlet-process prior,
as well as the two-step posterior inference procedure. We examine the
performance of our method through simulation studies in Section~\ref{sec:simulations}.
We apply our method to the time-course microarray gene expression from
the Notch experiment in Section~\ref{sec:cases}. Compared with
SplineCluster~[\citet{Heard.etal.2006}] and MCLUST~[\citet
{FraleyRaftery2002, FraleyRaftery2006}], our method appears to be more
accurate and sensitive
to subtle differences in different clusters, in both simulation studies
and the real application. In addition, the analysis of the real data reveals
several novel insights into the transcriptional dynamics after the
pulse of Notch activation.
We summarise and discuss the features of our method in
Section~\ref{sec:discussion}.\looseness=-1

\section{Random-effects mixture model}
\label{sec:model}
Consider $N$ genes measured at $J$ time points in each of the $R$ replicates.
Let $M_{ijr}$, $i=1, \ldots, N$, $j=1,\ldots, J$, $r=1, \ldots, R$, be
the measurement for the $i$th gene at the $j$th time point from the
$r$th replicate.
The $J$ time points may be unevenly distributed. We assume that there
are no missing data.
We use a random-effects mixture model to describe the heterogeneity in
replicated time-course data, and explain the details of the model below.

Following the standard mixture model formulation with a known number of
mixture components,
$K$, we assume that data vectors
\[
\M_i = (M_{i11}, \ldots, M_{i1R}, \ldots,
M_{iJ1}, \ldots, M_{iJR})^\tT
\]
are independent and identically distributed realizations drawn from a
mixture distribution with $K$ components and a set of mixing proportions
$w_k$, $k=1,\ldots, K$. The superscript $\tT$ represents ``transpose''
and ensures that $\M_i$ is a column vector. The probability density function
of $\M_i$, denoted by $f$, can be written as a weighted average:
\[
f (\M_i | \bTheta, \bSigma) = \sum^K_{k=1}
w_k g_k \bigl(\M_i | \bTheta^k,
\bSigma^k \bigr),
\]
where $g_k$ is the probability density function of the $k$th mixture
component, and $\bTheta=(\bTheta^1, \ldots, \bTheta^K)$ and $\bSigma
=(\bSigma^1, \ldots, \bSigma^K)$ are parameters of the mixture
distribution, with component-wise mean vector $\bTheta^k$ and
covariance matrix $\bSigma^k$, $k=1,\ldots,K$.
Whereas it is possible to define a cluster by more than one mixture
component, for presentation purposes we
consider here the case where one mixture component defines a cluster
and use ``mixture component'' and ``cluster''
interchangeably. Let $Z_i$ denote the cluster membership for the $i$th
gene. Then,
\[
\Pr(Z_i = k | \w, \bTheta, \bSigma) = w_k,
\]
where $\w$ is the set of mixing proportions.
Following the notation in \citet{Stephens2000a} and denoting the data by
$\M= (\M_1, \ldots, \M_N)$, we define the posterior allocation
probabilities as $\Pr(Z_i = k | \M)$,
$i=1,\ldots, N$ and $k=1,\ldots,K$, which form the posterior allocation
probability matrix $\P$ of dimension $N\times K$. We aim to estimate $\P$
as part of the inference and to form clusters based on the estimated $\P
$, using, for instance, the most likely allocation.

Inspired by variance components approaches~[\citet{Searle.etal.2006}]
and random-effects models frequently used in longitudinal studies~[\citet
{Dunson2010}], we constrain the covariance matrix
of each mixture component, $\bSigma^k$, by attributing the total
variability to three sources: clustering, sampling across multiple time
points (or more broadly speaking,
multiple experimental conditions), and sampling a limited number of replicates.
Whereas the first source of variability is due to ``grouping'' of the
genes, the latter two are defined by the design of the time-course experiment.
If the $i$th gene is sampled from the $k$th mixture component (i.e.,
$Z_i=k$), the random-effects model
can be written as follows:
%
%
\begin{equation}
\label{eqn:rem} M_{ijr} | \{Z_i=k\} =\Theta^k_{j}
+\phi^k_i+ \tau^k_{ij} +
\varepsilon^k_{ijr},
\end{equation}
where
\begin{eqnarray*}
\E \bigl(M_{ijr} | \{Z_i=k\} \bigr) &=&
\Theta^k_j,
\\
\phi^k_i | \bigl\{Z_i=k,
\lambda^k_\phi \bigr\} &\sim_\mathrm{i.i.d.}&
\mathrm{N} \bigl(0, \lambda^k_\phi \bigr),
\\
\tau^k_{ij} | \bigl\{Z_i=k,
\lambda^k_\tau \bigr\} &\sim_\mathrm{i.i.d.}&
\mathrm{N} \bigl(0, \lambda^k_\tau \bigr),
\\
\varepsilon^k_{ij} | \bigl\{Z_i=k,
\lambda^k_\varepsilon \bigr\} &\sim_\mathrm{i.i.d.}&
\mathrm{N} \bigl(0, \lambda^k_\varepsilon \bigr).
\end{eqnarray*}
In this formulation, $\Theta^k_j$ represents the ``true'' value (fixed
effect) at the $j$th time point, $\phi^k_i$ the within-cluster random
effect, $\tau^k_{ij}$ the
cross-experimental-condition random effect and $\varepsilon^k_{ijr}$ the
replicate random effect.
Here, the experimental conditions are time points.
We assume that random effects $\phi^k_i$, $\tau^k_{ij}$ and $\varepsilon
^k_{ijr}$ are independent across clusters and
of each other. Each of the three random effects has a corresponding
variability term:
$\lambda^k_\phi$ is the within-cluster variability, $\lambda^k_\tau$ the
cross-experimental-condition variability,
and $\lambda^k_\varepsilon$ the residual variability. The three types of variability are
all component specific.

Given cluster membership $Z_i=k$, replicated measurements of the $i$th
gene, $\M_i$, follow a multivariate normal distribution:
\[
\M_i | \bigl\{Z_i=k, \bTheta^k,
\lambda^k_\phi, \lambda^k_\tau,
\lambda^k_\varepsilon \bigr\} \sim_{\mathrm{ind}} \tN
_{JR} \bigl(\bTheta^k_{\mathrm{agg}}, \bSigma^k_{\mathrm{agg}}
\bigr),
\]
which has aggregated mean vector $\bTheta^k_{\mathrm{agg}} = ({\bTheta
^k}^\tT, \ldots, {\bTheta^k}^\tT)^\tT$, where $\bTheta^k$ repeats
$R$ times, and aggregated covariance matrix $\bSigma^k_{\mathrm{agg}}$
whose entry is
%
%
\begin{equation}
\label{eqn:multincov} \operatorname{Cov} (M_{ijr}, M_{ij'r'}) =
\lambda^k_\phi+ \lambda^k_\tau\one
\bigl(j=j' \bigr) + \lambda^k_\varepsilon \one
\bigl(j=j', r=r' \bigr),
\end{equation}
where $\one(A)$ is the indicator function that takes value 1 if
condition $A$ is satisfied and 0 otherwise.
In addition, $\M_i$ and $\M_j$, where $i\neq j$, are independent of
each other.

Parameters of interest under this random-effects mixture model include
the number of mixture components, $K$, component-specific parameters
$\bTheta^k$, $\lambda^k_\phi$, $\lambda^k_\tau$ and $\lambda^k_\varepsilon$, where $k=1, \ldots, K$, and
posterior allocation probability matrix $\P$ of dimension $N\times K$.

\section{Bayesian inference}
\label{sec:inference}
\subsection{The Dirichlet-process prior}
As mentioned in Section~\ref{sec:intro}, Dirichlet processes help to
partition the parameter space without prior knowledge of the number of
partitions, $K$,
and thus provide a coherent framework for directly estimating $K$ from
data and for sampling in a parameter space of variable dimensions.
Denote the parameter of interest for \textit{each} gene by $\bgamma_i$,
which, in our case, may include a mean vector $\bTheta_i$ and three
terms of variability,
namely, ${\lambda_\phi}_i$, ${\lambda_\tau}_i$ and ${\lambda_\varepsilon}_i$, such that
\begin{eqnarray*}
\bgamma_i &=& \{\bTheta_i, {\lambda_\phi}_i,
{\lambda_\tau}_i, {\lambda_\varepsilon}_i \},
\\
\M_i &\sim& F (\bgamma_i), \qquad i=1, \ldots, N,
\end{eqnarray*}
where $F$ represents a distribution, which is a multivariate normal
distribution in our case.
We assume that $\bgamma_i$s follow a random distribution $G$, which is
in turn a random draw from
a (compound) Dirichlet process, denoted as follows:
%
%
\begin{eqnarray}
\bgamma_i &\sim& G, \label{eqn:dir.def.1}
\\
G &\sim&\operatorname{DP} (\alpha, G_0), \qquad\alpha\geq0, \label
{eqn:dir.def.2}
\end{eqnarray}
with base distribution $G_0$ (continuous in our case), which describes
how values in the space are generated, and concentration parameter
$\alpha$,
which is nonnegative. Note that $\bgamma_i$s are identically
distributed, but not necessarily independent. The dependence among them
under the Dirichlet
process specifically refers to their values being clustered, that is,
some $\bgamma_i$s may take on identical value.

Indeed, the Dirichlet process describes a mechanism by which clustered
parameters $\bgamma_i$ may be simulated. We can generate a realization
for one of them, say, $\bgamma_1$, from $G_0$. The value of $\bgamma_2$
may be identical to $\bgamma_1$, with probability $1/(1+\alpha)$,
or an independent realization also from $G_0$ and different from
$\bgamma_1$, with probability $\alpha/(1+\alpha)$. Generally, having
generated $n$ realizations, the value
of the $n+1$st realization follows the following distribution~[\citet
{Antoniak1974}]:
%
%
\begin{eqnarray}
\label{eqn:dirichlet.sampling}&& \Pr(\bgamma_{n+1} = \bgamma|
\bgamma_1, \ldots, \bgamma_n, \alpha)
\nonumber
\\[-8pt]
\\[-8pt]
\nonumber
&&\qquad= \cases{
\displaystyle\frac{\sum^n_{i=1} \one(\bgamma_i = \bgamma) }{n+\alpha}, &\quad $\bgamma \in\{\bgamma_1,\ldots,
\bgamma_n\}$, \vspace*{2pt}
\cr
\displaystyle\frac{\alpha}{n+\alpha}, &\quad $\bgamma\notin \{
\bgamma_1,\ldots,\bgamma_n\}.$}
\end{eqnarray}
In other words, the probability of $\bgamma_{n+1}$
being identical to one of the existing values is proportional to the
number of times this value has already shown up. This sampling
process is also known as the Chinese restaurant process [reviewed in
\citet{Pitman2002}], a useful representation for \citet{Neal2000} to
derive the Metropolis--Hastings sampling procedures, which are explained
in the next section.

The sampling distribution above induces a distribution on the partition
of the~$N$ values, $\bgamma_1, \ldots, \bgamma_N$, with a random number
of partitions, $K$.
Specifically, the partition distribution is with respect to the cluster
memberships $Z_i$, $i=1,\ldots, N$, as well as $K$~[\citet{Antoniak1974}]:
%
%
\begin{equation}
\label{eqn:dir.prior} \Pr(Z_1, \ldots, Z_N, K| \alpha>0)
= \frac{\Gamma(\alpha)}{\Gamma
(\alpha+N)} \alpha^K \prod^K_{l=1}
(N_l-1)!,
\end{equation}
where $N_l$ is the size of the $l$th cluster, and
%
%
\begin{equation}
\label{eqn:dir.prior.a0} \Pr(Z_1= \cdots= Z_N, K=1|
\alpha=0) = 1.
\end{equation}
We use this distribution as the prior in our Bayesian inference.

As a measure of ``concentration,'' very small $\alpha$ leads to a small
probability of taking on a new value in the Dirichlet process, as
equation~(\ref{eqn:dirichlet.sampling}) suggests,
and hence to the probability mass being concentrated on a few distinct
values, as equations~(\ref{eqn:dir.prior}) and (\ref{eqn:dir.prior.a0})
suggest. As
$\alpha\rightarrow0$, $\bgamma_i$s are identical, which corresponds to
a single draw from the base distribution $G_0$. On the other hand, large
$\alpha$ leads to a large probability of taking on new values in the
Dirichlet process of equation~(\ref{eqn:dirichlet.sampling}) and an
appreciable probability
for having a range of distinct values in equation~(\ref{eqn:dir.prior}).
As $\alpha\rightarrow\infty$, $\bgamma_i$s are all different and form
an independent and identically distributed sample from $G_0$.
Therefore, $\alpha$ effectively controls the
sparsity of partitioning (or clustering).

We note that equations~(\ref{eqn:dirichlet.sampling})--(\ref
{eqn:dir.prior.a0}) characterise the canonical Dirichlet process with
parameter $\alpha$,
denoted DP$(\alpha)$, for an arbitrary space, as \citet{Antoniak1974}
defined it. The representation in expression~(\ref{eqn:dir.def.2}),
which we consider a compound Dirichlet process,
includes the additional information on how the elements of the space
arise: they are realizations of the base distribution $G_0$.

\subsection{A Metropolis--Hastings sampler for cluster memberships}
The key step in the MCMC algorithm is sampling partitions,
specifically, cluster memberships $Z_i$, under the Dirichlet-process
prior. We develop a Metropolis--Hastings sampler that allows
nonconjugate priors for parameters and efficient mixing.

Similar to \citet{Neal2000}, we design the MH procedure to sample each
$Z_i$ during an MCMC update. Let the current value of $Z_i$ be $z'$, which,
together with all the other $Z_j$, gives the current number of clusters
as $K=k'$. We propose a new value $z^*$ for $Z_i$, which gives rise to
the proposed value
$k^*$ for $K$. Let $\bxi$ be the parameter vector of interest for the
random-effects mixture model under the Dirichlet-process prior, such that
\[
\bxi\hspace*{-0.2pt}=\hspace*{-0.2pt} \bigl\{K, \bTheta_1, \ldots, \bTheta_K,
{\lambda^1_\phi}, \ldots, {\lambda^K_\phi}, {\lambda^1_\tau}, \ldots,
{\lambda^K_\tau}, {\lambda^1_\varepsilon}, \ldots, {\lambda^K_\varepsilon}, Z_1,
\ldots, Z_N, \alpha \bigr\}.
\]
We accept the proposal with probability $\operatorname{min}(1,H)$, where $H$ is the
Hastings ratio computed as follows:
%
%
\begin{eqnarray} \label{eqn:hastings.ratio}
H &=& \frac{\pi(Z_i=z^*)}{\pi(Z_i=z')} \frac
{g(Z_i=z'|Z_i=z^*)}{g(Z_i=z^*|Z_i=z')}
\nonumber\\
&=& \frac{\Pr(\M_i | Z_i=z^*, \cdot) \Pr(z_1, \ldots, z^*, \ldots, z_N,
k^*| \alpha)}{\Pr(\M_i | Z_i=z', \cdot) \Pr(z_1, \ldots, z', \ldots,
z_N, k'| \alpha)} \frac{g(z'|z^*)}{g(z^*|z')}
\\
&=& \frac{\Pr(\M_i | z^*, \cdot)}{\Pr(\M_i | z', \cdot)} \frac{\Pr
(z^*, k^*| \z_{-i}, \alpha)}{\Pr(z', k'| \z_{-i}, \alpha)} \frac
{g(z'|z^*)}{g(z^*|z')},
\nonumber
\end{eqnarray}
where $\cdot$ refers to current estimates of parameters in $\bxi$ other
than $Z_i$ and $\z_{-i}$ denotes the cluster memberships of all genes
except for the $i$th one,
which do not change when we update $Z_i$.

Under the Dirichlet-process prior, we can compute the conditional
probability $\Pr(z', k'| \z_{-i}, \alpha)$ as in Proposition~\ref
{prop:condprob}:
%
\begin{prop}
\label{prop:condprob}
Consider $N$ values drawn from a Dirichlet process with concentration
parameter $\alpha\geq0$. These values can be partitioned into
$K$ clusters, where $K$ is a random variable, with $Z_i$, $i=1,\ldots,
N$, indicating the cluster membership. Then the following conditional
probability holds:
%
%
\begin{eqnarray}
\label{eqn:dirichlet.cond} &&\Pr(Z_i=z, K=k| \Z_{-i} =
\z_{-i}, \alpha)
\nonumber
\\[-8pt]
\\[-8pt]
\nonumber
&&\qquad= \cases{\displaystyle \frac{N_{z} -1}{N-1+\alpha}, &\quad $\mbox{$Z_i$
is not in a singleton cluster} ,$\vspace*{2pt}
\cr
\displaystyle\frac{\alpha}{N-1+\alpha}, & \quad$
\mbox{$Z_i$ is in a singleton cluster},$}
\end{eqnarray}
where $\Z_{-i}$ with value $\z_{-i}$ denotes the cluster memberships,
excluding the $i$th gene, and $N_z$ is the size of the $z$th cluster.
\end{prop}
\begin{pf}
See the \hyperref[sec:app]{Appendix}.
\end{pf}

\citet{Neal2000} then proposed an MH procedure, using the conditional
probability in equation~(\ref{eqn:dirichlet.cond}) as the proposal
distribution $g$, which led
to a simplified Hastings ratio:
%
%
\begin{equation}
\label{eqn:h.neal} H = \frac{\Pr(\M_i | z^*, \cdot)}{\Pr(\M_i | z', \cdot)}.
\end{equation}
The main problem with this MH sampler is slow mixing: because the
probability of a move is proportional to the size of the cluster, the
Markov chain
can be easily stuck, especially when there exist one or a few large
clusters. For example, consider $N=200$ and current clusters 1--3 of
size 185, 10 and 5, respectively.
A~gene currently allocated to cluster 1 may be much more similar to
cluster 3, implying a high likelihood ratio as in the simplified
Hastings ratio (\ref{eqn:h.neal}).
However, the probability of proposing such a favorable move from
cluster 1 to cluster 3 is only $5/(199+\alpha)$,
where $\alpha$ is usually small to induce a parsimonious
partition. The probability of moving a gene to a previously nonexistent
cluster is $\alpha/(199+\alpha)$, which can be even
smaller.\looseness=1

We develop a novel MH MCMC strategy to deal with poor mixing of Neal's
MH sampler. Our proposal distribution for a cluster membership is discrete
uniform on the integer set from 1 to $k'+1$, excluding the current
cluster the gene belongs to, where $k'$ is the number of existing
clusters. This proposal
distribution forces the proposed cluster membership always to be
different from the current one, and makes the Markov chain move to a
new or small cluster more easily. Whether to accept the proposal or not
depends on the
Hastings ratio, which needs to be recalculated as in Proposition~\ref
{prop:hastings}.
%
\begin{prop}
\label{prop:hastings}
For cluster membership $Z_i$ with current value $z'$, if proposal $z^*$
is generated from a discrete uniform distribution
over the integer set $\{1, \ldots, z'-1, z'+1, \ldots, k'+1\}$,
where $k'$ is the current number of clusters, then the Hastings ratio
takes on values as listed in Table~\ref{table:hastings}, where
four cases, including a generation of a new cluster and elimination of
an existing cluster, are considered.
\end{prop}
\begin{pf}
The proof of this proposition can be found in Section~1 of the
supplemental material [\citet{Fu.etal.2013.supp}].
\end{pf}
%
\begin{table}
\caption{Hastings ratio for four cases under the proposed
Metropolis--Hastings sampler for cluster membership $Z_i$ with current
value $z'$ and proposed
value $z^*$. $k^*$ and $k'$ are the number of clusters after and
before the proposed move, respectively}\label{table:hastings}
\begin{tabular*}{\textwidth}{@{\extracolsep{\fill}}lcccc@{}}
\hline
& \textbf{Current cluster} & \textbf{Proposal} & & \\
& \textbf{a singleton} &\textbf{an existing label} & $\bolds{k^*-k'}$ & \multicolumn{1}{c@{}}{\textbf{Hastings ratio}} \\
\hline
1 & Yes & Yes & $-1$ & $ \frac{\Pr(\M_i | z^*, \cdot
)}{\Pr(\M_i | z', \cdot)} \frac{N_{z^*}}{\alpha} \frac{k'}{k'-1}$\\[3pt]
2 & Yes & No & 0 & $ \frac{\Pr(\M_i | z^*, \cdot)}{\Pr
(\M_i | z', \cdot)} $\\[3pt]
3 & No & Yes & 0 & $ \frac{\Pr(\M_i | z^*, \cdot)}{\Pr
(\M_i | z', \cdot)} \frac{N_{z^*}}{N_{z'}-1}$\\[3pt]
4 & No & No & 1 & $ \frac{\Pr(\M_i | z^*, \cdot)}{\Pr(\M
_i | z', \cdot)} \frac{\alpha}{N_{z'}-1} \frac{k'}{k'+1}$\\
\hline
\end{tabular*}
\end{table}

\subsection{Other prior distributions}
\label{sec:priors}
The base distribution $G_0$ specifies the prior on the cluster mean
vector $\bTheta_k$, each of the three types of
variability ${\lambda^k_\phi}$, ${\lambda^k_\tau}$ and ${\lambda^k_\varepsilon}$, for all $k$. We use a uniform
distribution on $[0, u]$ as the prior for the $\lambda$s,
and experiment with different values of the upper bound $u$. Values of
$u$ are guided by the data.

We experiment with three options for $\bTheta_k$: (i) a zero vector
of length~$J$, where $J$ is the number of time points. This is a
natural choice for our data considering that the relative gene
expression level on the $\log_2$ scale
is not much different from~0; (ii) a realization generated from an
Ornstein--Uhlenbeck (OU) process~[\citet{Merton1971}].
An OU process has four parameters: the starting value, the mean and
variation of the process, and the
mean-reverting rate. We therefore specify the normal distribution of
the starting value, and the normal distribution of the process mean,
the uniform distribution of the process variation, and the gamma
distribution for the mean-reverting rate; and (iii) a
realization generated from a Brownian motion with drift. This process
has three parameters: the starting value, the mean and
the variation~[\citet{TaylorKarlin1998}]. Similarly, we specify the
normal distribution of the starting value, and the normal distribution
of the
process mean, and the uniform distribution of the process variation.
Values of the parameters in these distributions are
again guided by the summary statistics of the data.

For the concentration parameter $\alpha$, we experiment with two
options: (i)~a~Gamma prior with the shape and rate parameters,
which can be updated by a Gibbs sampler, as described in \citet
{EscobarWest1995}; and (ii)~a~uniform prior on $[0,u']$, where
$u'$ can be different values, which is updated by an MH sampler [see
Section~2 of the supplemental material; \citet{Fu.etal.2013.supp}].

\subsection{\texorpdfstring{The MCMC algorithm for $\bxi$}{The MCMC algorithm for xi}}
The complete MCMC algorithm for sampling $\bxi$ consists of two major
steps in each iteration:
\begin{longlist}[\textit{Step} 1.]
\item[\textit{Step} 1.] For each $i$ from 1 to $N$, update $Z_i$ using
the MH sampler described above;
\item[\textit{Step} 2.] Given the partition from \textit{Step 1}, update
other parameters in $\bxi$ using Gibbs or MH samplers. Details of this
step are in Section~2 of the supplemental material [\citet{Fu.etal.2013.supp}].
\end{longlist}

If the total number of MCMC iterations is $S$, then the time complexity
of this MCMC algorithm is roughly $\bigo(SJR(4N+K))$, where 4 comes
from the steps
required in the MH sampler described above, such as generating a
proposal, computing the likelihoods and the Hastings ratio.

\subsection{Two-step posterior inference under the Dirichlet-process prior}
For probabilistic clustering, we would like to estimate the posterior
allocation probability matrix $\P$ of dimension $N\times K$
with entries $p_{ik} = \Pr(Z_i = k | \M)$, each of which is the
probability of the $i$th gene belonging to the $k$th cluster given the data.
This matrix is not part of the parameter vector $\bxi$ and is therefore
not sampled during MCMC. Below, we propose resampling
followed by relabeling to estimate $\P$ from $H$ MCMC samples of $\bxi
$, while dealing with label-switching~[\citet{Stephens2000b}]:
\begin{longlist}[1.]
\item[1.]\textit{Resampling}: Let $\Q^{(h)}$ of dimension $N\times K^{(h)}$,
whose entries are $q^{(h)}_{ik}$, $h=1,\ldots, H$, be the posterior
allocation probability matrix
from the $h$th MCMC sample with arbitrary labeling.
The resampling step builds upon an alternative representation of the
Dirichlet process as an infinite mixture model~[\citet{Neal2000, Green2010}].
Specifically, for a Dirichlet process defined in equations~(\ref
{eqn:dir.def.1}) and~(\ref{eqn:dir.def.2}) with concentration parameter
$\alpha$ and base distribution $G_0$,
an infinite mixture model representation corresponds to taking the
limit in the finite mixture model below, letting $K\rightarrow\infty$
and $\alpha^*\rightarrow0$, such that
$\alpha^* K \rightarrow\alpha$~[\citet{Green2010}]:
\begin{eqnarray*}
\bgamma^*_k &\sim& G_0,\qquad k=1,\ldots, K,
\\
(w_1, \ldots, w_K) &\sim&\operatorname{Dirichlet}_K
\bigl(\alpha^*, \ldots, \alpha^* \bigr),
\\
\Pr \bigl(\bgamma_i=\bgamma^*_k \bigr) &=&
\Pr(Z_i = k) = w_k.
\end{eqnarray*}
Conditional on the $h$th MCMC sample, the mixture model for the data
becomes finite:
\begin{eqnarray*}
\bigl(w^{(h)}_1,\ldots, w^{(h)}_{K^{(h)}} \bigr)
&\sim&\operatorname {Dirichlet}_{K^{(h)}} \bigl(\alpha^{(h)}, \ldots,
\alpha^{(h)} \bigr),
\\
\Pr \bigl(Z^{(h)}_i = k | \w^{(h)} \bigr) &=&
w^{(h)}_k,
\\
\M_i | \bigl\{Z^{(h)}_i=k,
\bTheta^{(h)}_{k}, \bSigma^{(h)}_k \bigr\}
& \sim& N_{JR} \bigl(\bTheta^{(h)}_{k},
\bSigma^{(h)}_k \bigr).
\end{eqnarray*}
Then, the posterior probability $q^{(h)}_{ik}$ can be sampled from the
following distribution using the $h$th MCMC sample $\bxi^{(h)}$:
\begin{eqnarray*}
q^{(h)}_{ik} &=& \Pr \bigl(Z^{(h)}_i = k |
\M, \bxi^{(h)} \bigr)
\\
&\propto& N_{JR} \bigl(\M_i|\bTheta^{(h)}_{k},
\bSigma^{(h)}_k \bigr) w^{(h)}_k
\\
&\propto& N_{JR} \bigl(\M_i|\bTheta^{(h)}_{k},
\bSigma^{(h)}_k \bigr) \operatorname {Dirichlet}_{K^{(h)}}
\bigl(w^{(h)}_k| \alpha^{(h)}, \ldots,
\alpha^{(h)} \bigr),
\end{eqnarray*}
where mixing proportion $w^{(h)}_k$ is generated from a (conditionally)
finite Dirichlet distribution. The time complexity of this step is
roughly\break $\bigo(H(NJR+K))$.

\item[2.]\textit{Relabeling}: Labels in $\Q^{(h)}$, $h=1,\ldots,H$, of
dimension $N\times K^{(h)}$, are arbitrary: for example, cluster \#2 in
$\Q^{(s)}$
does not necessarily correspond to cluster \#2 in $\Q^{(t)}$, where
$s\neq t$.
To deal with arbitrary labeling (also known as ``label-switching''), we
apply the relabeling algorithm from \citet{Stephens2000b} (Algorithm~2
in that paper)
to matrices $\Q$ to ``match'' the labels across MCMC samples. The
dimension of $\Q$s are set to be $N\times K_{\max}$, where $K_{\max}$
is the
maximum number of clusters from all recorded MCMC samples. We fill in
matrices of lower dimensions with 0s such that all $\Q$s have the same
dimension.
Stephens' relabeling algorithm then finds a set of permutations, one
for the columns of each $\Q$, and the resulting matrix $\P$, such that
the Kullback--Leibler
distance between $\P$ and column-permuted $\Q$s is minimised. Details
of our application, which also
implements the Hungarian algorithm [aka Munkres assignment algorithm;
\citet{Kuhn1955, Munkres1957}] for minimisation can be found in Section~3 of the
supplemental material [\citet{Fu.etal.2013.supp}]. If $L$ is the number
of iterations for the relabeling step to achieve convergence, then the
time complexity of this step is roughly
$\bigo(LH(NJR+K^3))$, as the time complexity of the Hungarian algorithm
is $\bigo(K^3)$~[\citet{Munkres1957}].
\end{longlist}

\section{Simulations}
\label{sec:simulations}
We investigate the performance of our MH MCMC algorithm and compare the
clustering performance of our method with MCLUST [\citet{FraleyRaftery2006}]
and SplineCluster [\citet{Heard.etal.2006}] on data sets simulated from
multiple settings, each with different values of variabilities.
The size of each data set is comparable to the number of differentially
expressed genes we identify from the real time-course data, which we
introduced in Section~\ref{sec:intro}
and will describe in detail in Section~\ref{sec:cases}: the number of
items $N$ is between 100 and 200, the number of
experimental conditions (time points) $J$ is 18, and the number of
replicates $R$ is 4. The last two values are identical to those of the
real data. However,
to keep track of the parameters for individual clusters, we consider
only 6 clusters instead of the 14 or 19 clusters our method infers for
the real data (Section~\ref{sec:cases}).

For each cluster,
we simulated data from a multivariate normal distribution.
Specifically, we generated the mean vector from an Ornstein--Uhlenbeck
(OU) process with three parameters,
which are the initial value, the overall mean and the mean-reverting
rate. We constructed the covariance matrix as specified in equation~(\ref{eqn:multincov}) with true
values of the three types of variability (Table~\ref{table:simu.data}).
In simulations~\#1 and \#2, all three types of variability are nonzero,
with simulation \#2
having more extreme within-cluster variability in some clusters. In
particular, the level of different types of variability in simulation~\#
1 is largely comparable to that of
6 of the 14 clusters our method infers for the real time-course data
(Section~\ref{sec:cases}). In simulations \#3 and \#4, only the
residual variability is nonzero,
with simulation \#4 having high variability in some clusters. The
simplified covariance structure in the latter two simulations has been
adopted in SplineCluster and
other methods~[\citet{MedvedovicSivaganesan2002, Medvedovic.etal.2004,
Qin2006}]. Since SplineCluster and MCLUST allow only one replicate per
item, we average over the
replicates in simulated data and use these sample means as input for
SplineCluster and MCLUST, and use default settings in both programs.
Also note that neither
DIRECT nor MCLUST assumes temporal dependence, whereas SplineCluster does.

\begin{table}
\caption{Key parameter values used in four sets of simulations. Ten
data sets were simulated under each setting.
The true number of clusters is 6. True standard deviations of the three
types of variability (within-cluster variability,
cross-experimental-condition variability and residual variability) are given.
Size refers to the number of items simulated for each cluster. Standard
deviations used in simulation \#1 are close to some of the clusters inferred
for the real time-course data}\label{table:simu.data}
\begin{tabular*}{\textwidth}{@{\extracolsep{\fill}}lcd{1.2}d{1.2}d{1.2}c@{}}
\hline
& & \multicolumn{3}{c}{\textbf{Standard deviation}} & \\[-6pt]
& & \multicolumn{3}{c}{\hrulefill} & \\
& & \multicolumn{1}{c}{\textbf{Within-cluster}} & \multicolumn{1}{c}{\textbf{Expt. cond.}} & \multicolumn{1}{c}{\textbf{Resid.}} & \\
\multicolumn{1}{@{}l}{\textbf{Simulations (reps)}} & \multicolumn{1}{c}{$\bolds{K}$} & \multicolumn{1}{c}{$\bolds{\sqrt{{\lambda_\phi}}}$} &
\multicolumn{1}{c}{$\bolds{\sqrt{{\lambda_\tau}}}$} & \multicolumn{1}{c}{$\bolds{\sqrt{{\lambda_\varepsilon}}}$} & \multicolumn{1}{c@{}}{\textbf{Size}}
\\
\hline
\#1 (10) & 6 & 0.05 & 0.01 & 0.2 & 80 \\
& & 0.1 & 0.05 & 0.2 & 20 \\
& & 0.1 & 0.05 & 0.2 & 10 \\
& & 0.1 & 0.05 & 0.1 & 10 \\
& & 0.2 & 0.1 & 0.2 & 70 \\
& & 0.5 & 0.1 & 0.6 & 10 \\[3pt]
\#2 (10) & 6 & 0.01 & 0.5 & 0.5 & 20 \\
& & 0.1 & 0.5 & 0.5 & 20 \\
& & 0.1 & 0.5 & 0.5 & 20 \\
& & 0.5 & 0.5 & 0.5 & 20 \\
& & 0.5 & 0.5 & 0.5 & 20 \\
& & 1 & 0.5 & 0.5 & 20 \\[3pt]
\#3 (10) & 6 & 0 & 0 & 0.26 & 80 \\
& & 0 & 0 & 0.35 & 20 \\
& & 0 & 0 & 0.35 & 10 \\
& & 0 & 0 & 0.25 & 10 \\
& & 0 & 0 & 0.50 & 70 \\
& & 0 & 0 & 1.20 & 10 \\[3pt]
\#4 (10) & 6 & 0 & 0 & 1.01 & 20 \\
& & 0 & 0 & 1.1 & 20 \\
& & 0 & 0 & 1.1 & 20 \\
& & 0 & 0 & 1.5 & 20 \\
& & 0 & 0 & 1.5 & 20 \\
& & 0 & 0 & 2.0 & 20 \\
\hline
\end{tabular*}
\end{table}

\begin{table}
\caption{Comparison of methods on simulated data in terms of the
corrected Rand Index~[\citet{HubertArabie1985}] to assess
clustering accuracy: the higher the corrected Rand Index, the closer
the inferred clustering is to the truth. Each cell displays the mean (standard
deviation in parentheses) of the Rand Index
over the 10 data sets simulated under each setting. Highest values
(accounting for variability) in each scenario are highlighted}
\label{table:simu.compare}
\begin{tabular*}{\textwidth}{@{\extracolsep{\fill}}lcccc@{}}
\hline
\multicolumn{1}{@{}l}{\textbf{Simulations}} & \multicolumn{1}{c}{\textbf{True} $\bolds{K}$} & \multicolumn{1}{c}{\textbf{DIRECT}} & \multicolumn{1}{c}{\textbf{SplineCluster}} &
\multicolumn{1}{c@{}}{\textbf{MCLUST}} \\
\hline
\#1 & 6 & \textbf{0.99 (0.01)} & 0.84 (0.02) & 0.60 (0.13) \\
\#2 & 6 & \textbf{0.69 (0.08)} & 0.47 (0.10) & \textbf{0.71 (0.06)} \\
\#3 & 6 & \textbf{0.99 (0.01)} & \textbf{1.00 (0.00)} & \textbf{1.00 (0.00)} \\
\#4 & 6 & \textbf{0.95 (0.04)} & 0.47 (0.00) & \textbf{0.97 (0.03)} \\
\hline
\end{tabular*}
\end{table}

\begin{table}
\tabcolsep=4pt
\caption{Comparison of methods on simulated data in terms of the number
of nonsingleton (NS)
clusters and the number of singleton (S) clusters inferred under each
method. Each cell displays
the mean (standard deviation in parentheses) number of clusters over
the 10 data sets simulated
under each setting. The NS number closest to the truth (i.e., 6) in
each scenario is highlighted}\label{table:simu.compare.2}
\begin{tabular*}{\textwidth}{@{\extracolsep{\fill}}lccccccc@{}}
\hline
& & \multicolumn{2}{c}{\textbf{DIRECT}} & \multicolumn{2}{c}{\textbf{SplineCluster}} &
\multicolumn{2}{c@{}}{\textbf{MCLUST}} \\[-6pt]
& & \multicolumn{2}{c}{\hrulefill} & \multicolumn{2}{c}{\hrulefill} &
\multicolumn{2}{c@{}}{\hrulefill} \\
\textbf{Simulations} & \multicolumn{1}{c}{\textbf{True} $\bolds{K}$} & \multicolumn{1}{c}{\textbf{NS}} & \multicolumn{1}{c}{\textbf{S}} &
\multicolumn{1}{c}{\textbf{NS}} & \multicolumn{1}{c}{\textbf{S}} & \multicolumn{1}{c}{\textbf{NS}} & \multicolumn{1}{c@{}}{\textbf{S}} \\
\hline
\#1 & 6 & \textbf{6.2 (0.4)} & \phantom{0}1.7 (1.1) & 7.3 (0.5) & 0.0 (0.0) & 12.0
(2.2) & 0.0 (0.0) \\
\#2 & 6 & \textbf{7.5 (1.4)} & 19.6 (7.2) & 3.8 (0.6) & 0.2 (0.4) & \phantom{0}7.7
(1.1) & 0.1 (0.3) \\
\#3 & 6 & 6.2 (0.6) & \phantom{0}0.6 (0.5) & \textbf{6.0 (0.0)} & 0.0 (0.0) & \phantom{0}\textbf{
6.0 (0.0)} & 0.0 (0.0) \\
\#4 & 6 & 6.1 (0.3) & \phantom{0}2.8 (2.2) & 3.0 (0.0) & 0.0 (0.0) & \phantom{0}\textbf{6.0
(0.0)} & 0.0 (0.0) \\
\hline
\end{tabular*}
\end{table}

Table~\ref{table:simu.compare} summarises the performance of DIRECT and
compares it to that of SplineCluster and MCLUST. Correctly inferring
the number of clusters is key to the overall performance: when the
inferred number of clusters is close to the truth, all three
methods manage to allocate most of the items to the right clusters and
thus achieve a high corrected Rand Index, and vice versa (Tables~\ref{table:simu.compare}
and~\ref{table:simu.compare.2}). Below we discuss the performance of
each method in turn.\looseness=-1

DIRECT recovers the true clustering consistently well in all the
simulations, obtaining high accuracy of cluster assignments of individual
items, which is reflected in the high corrected Rand Index (Table~\ref{table:simu.compare}). Accuracy and consistency come from
recovering the true number of (nonsingleton) clusters, as indicated in
Table~\ref{table:simu.compare.2}. This good performance persists even
when the data were simulated
under the ``wrong'' model (simulations \#3 and \#4). However, DIRECT
tends to produce singleton clusters, when those singletons are
simulated from clusters of high variation (Table~\ref{table:simu.compare.2}).

MCLUST achieves high accuracy in three out of the four simulations.
However, its performance
is much worse than DIRECT and SplineCluster in simulation \#1: MCLUST
tends to infer a higher number of clusters with large variability
(Table~\ref{table:simu.compare.2}).\vadjust{\goodbreak}

In contrast, SplineCluster tends to infer fewer clusters for more
heterogeneous data. The dependence structure in simulations \#3 and \#4
is in fact employed in SplineCluster. However, while SplineCluster
infers the
number of clusters correctly and allocates the items correctly in
simulation \#3, it infers a much lower number of clusters in simulation
\#4, which leads to a
much lower corrected Rand Index (Tables~\ref{table:simu.compare}
and~\ref{table:simu.compare.2}).
The heterogeneity in simulation \#4 (as well as in simulation \#2) is
too high for SplineCluster to distinguish different clusters, it therefore
settles on a more parsimonious clustering than the truth.

\section{Application to time-course gene expression data}
\label{sec:cases}
\subsection{Experimental design and data preprocessing}
As explained in the \hyperref[sec:intro]{Introduction}, gene expression data were collected
using two-colour microarrays from four independent biological replicates
of \textit{Drosophila} adult muscle cells at 18 unevenly-spaced time points
(in minutes): 0, 5, 10, 15, 20, 25, 30, 35, 40, 50, 60, 70, 80, 90,
100, 110, 120, 150, where 0 is the start of a 5-minute treatment of
Notch activation~[\citet{Housden2011}].
Similar to other gene expression data, the expression measured here is
in fact the relative expression of treated cells to control cells,
evaluated as the $\LOG2$ fold change.
The two colours of the microarray were used to distinguish treated and
control cells. We applied quantile normalization to the distributions
of spot intensities of the two colours
across all $18\times4=72$ arrays. Mapping of the oligonucleotide probes
on the microarray to the \textit{Drosophila} genome followed FlyBase
release 4 and
earlier for \textit{Drosophila melanogaster}. After the initial quality screen
we retained 7467 expressed genes, that is, the absolute
expression
levels of genes in the treated and control cells are detectable by the
microarray. These retained genes
are about half of the \textit{Drosophila} genome. We further imputed
missing values in the temporal profiles of these genes [see Section~4
of the supplemental material;
\citet{Fu.etal.2013.supp}].
These data were challenging to analyse, as the (relative) expression
levels of most of these genes were close to~0.
To identify differentially expressed (DE) genes over the time course, we
applied EDGE~[\citet{Storey.etal.2005}] to identify 163 such genes at a
false discovery rate (FDR) of 10$\%$ and 270 genes at an FDR of 25$\%$.
However, even
among the 163 DE genes, the (relative) expression levels are generally
very low (Figure~\ref{fig:de163.mean}).

\subsection{Results from DIRECT}
We ran DIRECT multiple times on both data sets with
different initial values.
Each run consisted of 10,800 iterations, including 20\% burn-in.
MCMC samples were recorded every 54th iteration. These runs each took
about 8 hours for 163 genes and 12 hours for 270 genes on 2.3 GHz CPUs,
including
approximately 1 hour for resampling and a few minutes for relabeling.
Since the results were consistent across runs, we report below the
results from only one run for each data set, averaging the inferred
posterior allocation probability
matrix across MCMC iterations and defining clusters in terms of the
most likely allocations a posteriori.

\begin{figure}

\includegraphics{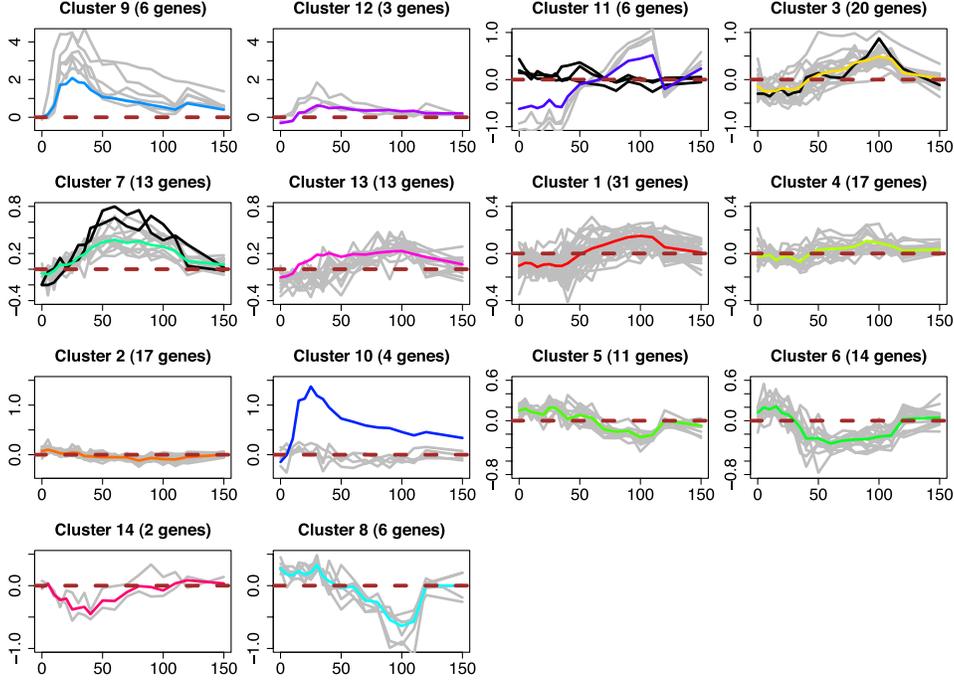}

\caption{Mean profiles (gray and black lines) of individual genes in
clusters inferred under our DIRECT method for the 163 significantly
expressed genes.
Each pair of plots, starting from the top left panel, display the same
range on the vertical axis. Each coloured line is the posterior mean
estimate of the cluster-specific mean vector.
Because more genes than those allocated to a cluster may have been used
for inference of the mean vector,
the coloured curves (inferred mean vectors) are not necessarily located
amid the profiles of the genes in that cluster (e.g., cluster \#10,
which shows a
rather extreme example). Genes with black lines
are analysed in more detail and presented in Figure~\protect\ref
{fig:de163.3genes}. In particular, the three genes with black lines in
cluster \#11 are
also allocated to cluster \#10 or cluster \#5 with a similar posterior
probability (see Figure~\protect\ref{fig:de163.3genes}).}
\label{fig:de163.curves}
\end{figure}

\begin{figure}

\includegraphics{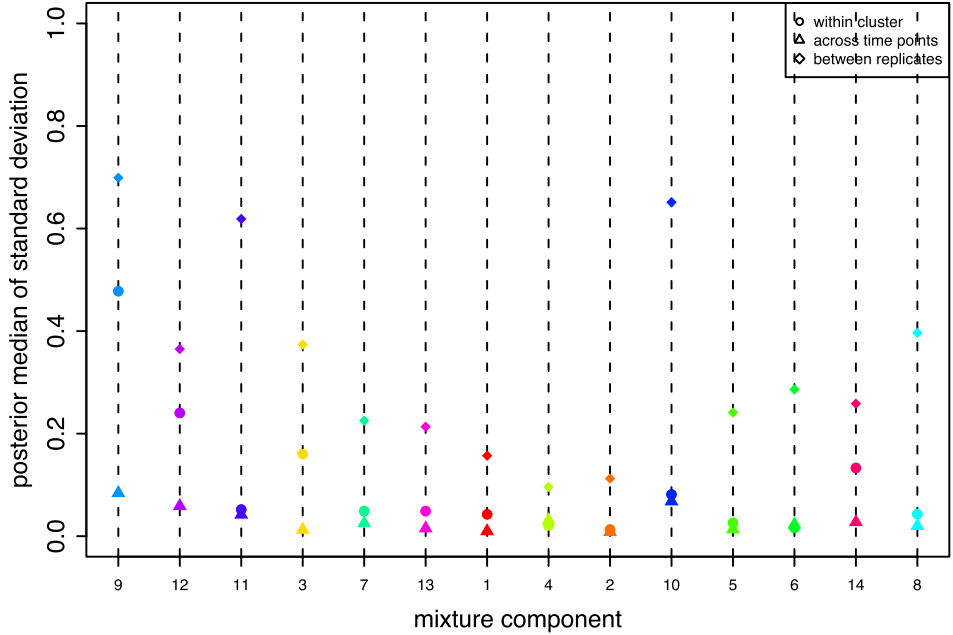}

\caption{Posterior median estimates of standard deviations from our
DIRECT program for the three types of variability in each inferred
mixture component
for the 163 significantly expressed genes.
Colours and numbering match those in Figure~\protect\ref{fig:de163.curves}.}
\label{fig:de163.var}
\end{figure}

Our DIRECT method identified 14 clusters for the 163 genes. Clusters
differ in both the mean vectors (Figure~\ref{fig:de163.curves}) and the
three types of
variability (Figure~\ref{fig:de163.var}). The cluster means differ in
the magnitude and timing of the maximal or minimal expression. Because
more genes
than those allocated to a cluster may have been used for inference of
the mean vector, the inferred mean vectors (represented by the coloured
curves) are not necessarily
located amid the profiles of the genes in that cluster (e.g., cluster \#
10, which shows a rather extreme example).
In terms of variability, the inferred clusters are homogeneous visually
and numerically: the within-cluster variability is small for most
inferred clusters, whereas in all clusters the majority
of the variability left unexplained by the mixture model is the
residual variability, which is the variability between replicates
(Figure~\ref{fig:de163.var}). In several
clusters, such as \#9, \#12 and \#14, the estimated within-cluster
variability in Figure~\ref{fig:de163.var} may seem higher than the
clustered mean profiles would indicate
(Figure~\ref{fig:de163.curves}). This is because, as mentioned earlier,
our probabilistic clustering method estimated these variability terms
using more genes than those assigned to the
corresponding cluster based on the highest posterior allocation
probability. Including these additional genes may increase the
within-cluster variability.

\begin{figure}

\includegraphics{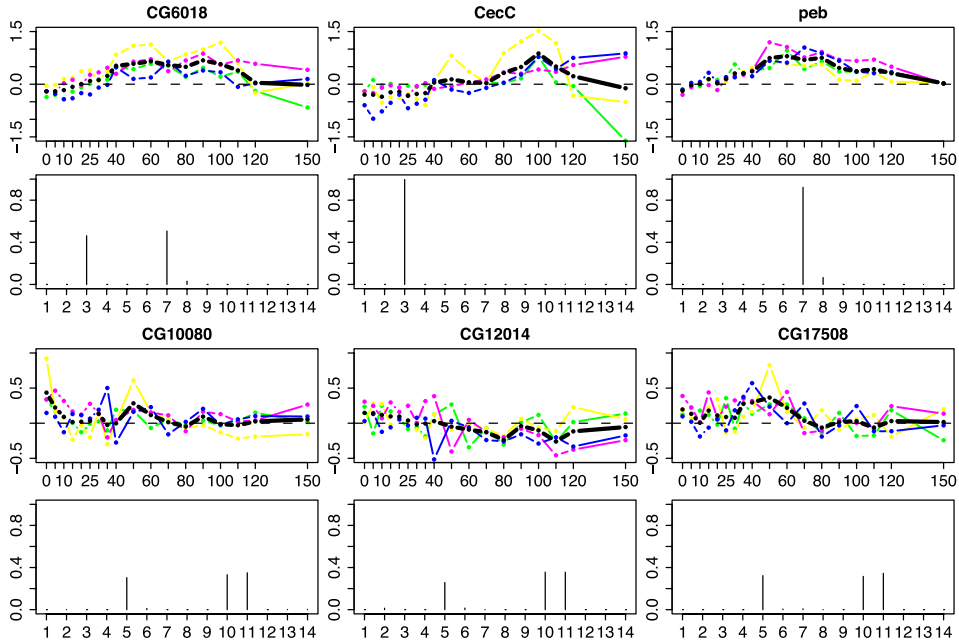}

\caption{Replicated and mean temporal profiles, as well as posterior
allocation probabilities, of six genes from the 163 gene set.
These genes correspond to
the black lines in Figure~\protect\ref{fig:de163.curves}. For each
gene, the top plot
shows the replicated (coloured) and mean (black) temporal profiles.
Colouring here indicates replicates rather than clustering.
The bottom plot shows the inferred
posterior probabilities (vertical lines) of allocating the
corresponding gene to a cluster (or mixture component).
The lengths of the vertical lines sum up to 1 in each of these three plots.}
\label{fig:de163.3genes}
\end{figure}

\begin{figure}

\includegraphics{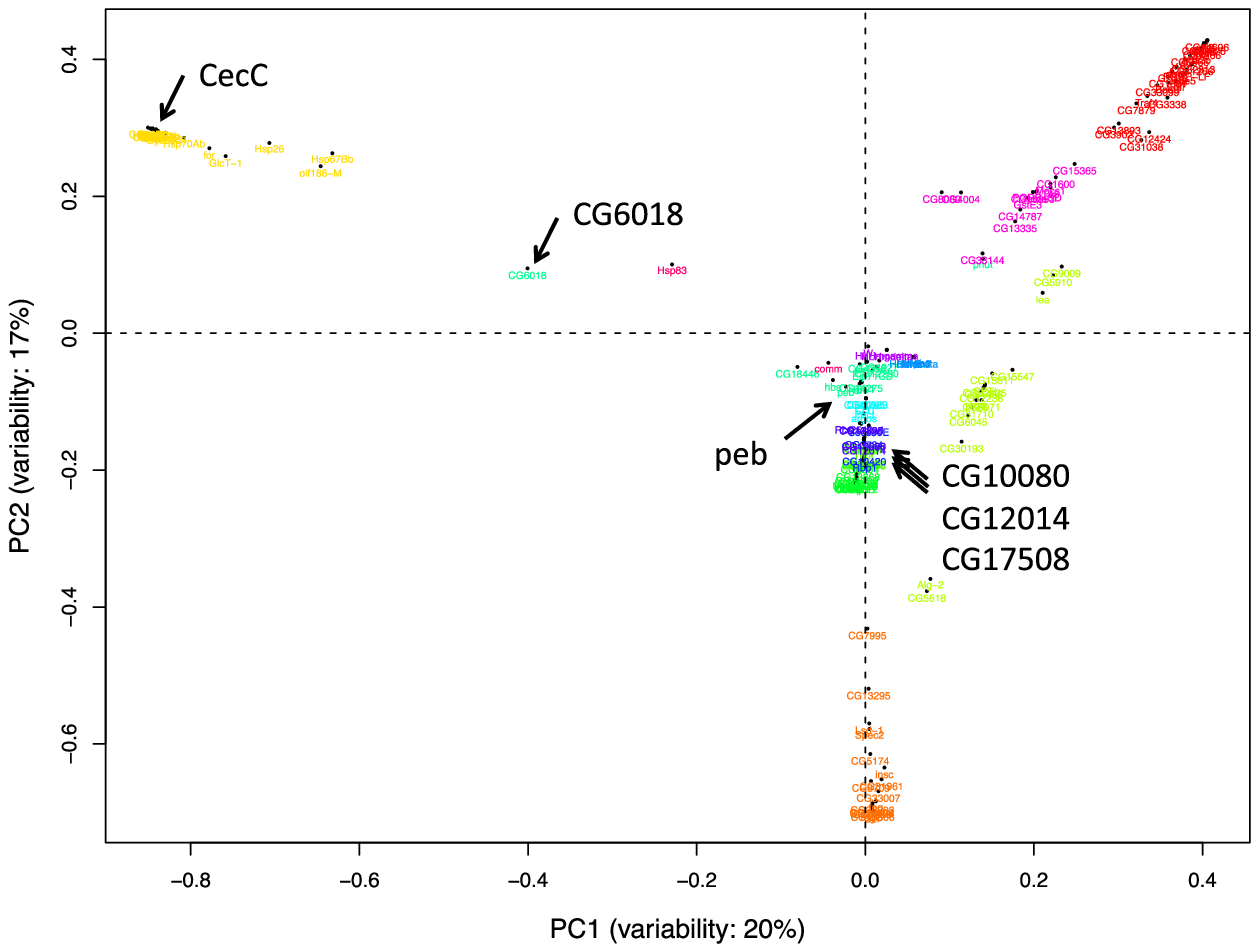}

\caption{PCA plot of the posterior allocation probability matrix for
163 genes. These colours match those in Figures~\protect\ref
{fig:de163.curves} and
\protect\ref{fig:de163.var}. Six arrows point to the six genes also
highlighted in Figure~\protect\ref{fig:de163.curves} and examined in
Figure~\protect\ref{fig:de163.3genes}.}
\label{fig:de163.pca}
\end{figure}

Whereas the mean profile plot (Figure~\ref{fig:de163.curves}) and the
variability plot (Figure~\ref{fig:de163.var}) visualise different
features of inferred clusters, they do
not display the uncertainty and complexity in inferred clustering. For
example, gene \textit{CG6018}, inferred to belong to cluster \#3 (with
peak expression appearing
around 100 min; very late response) with probability 0.51, also has a
substantial\vadjust{\goodbreak} probability of 0.46 to be associated with cluster \#7 (with
peak expression appearing
between 50 and 100 min; late response); see Figure~\ref{fig:de163.3genes}. Indeed, the replicated profiles of this gene show
similarity to the cluster mean profiles
of both clusters. Our inference indicates that the temporal profile of
\textit{CG6018} is better described by a two-component mixture
distribution, sharing features with both clusters. In contrast, the
profiles of genes \textit{Cecropin C} (or \textit{CecC}) and \textit{pebbled}
(or \textit{peb}) can be
adequately represented by one multivariate normal component (Figure~\ref{fig:de163.3genes}). Three genes, \textit{CG10080}, \textit{CG12014} and \textit{CG17508}, are
better described by a three-component mixture distribution, that is,
their expression profiles share features with three clusters
(Figure~\ref{fig:de163.3genes}).

We apply
principal components analysis (PCA) to the posterior allocation
probability matrix to visualise the uncertainty and complexity in
clustering. Figure~\ref{fig:de163.pca}
shows the scores of the probability matrix based on the first two
principal components. Since each row of the probability matrix
represents the distribution of cluster allocation
for an individual gene, the PCA plot displays the positions of
individual genes relative to their allocated clusters and to other
clusters. Genes with similar posterior allocation probabilities are
located next to each other. Specifically, most of the genes are
allocated to a single cluster with probability
above 0.8 and stay close to each other in the same cluster on the PCA
plot. On the other hand, genes associated with multiple clusters each
with a substantial probability
are located in between those clusters. For example, the aforementioned
gene \textit{CG6018} is positioned between clusters \#3 and \#7 on this plot.

To examine the sensitivity of our method to specification of the
priors, we experimented with different options regarding the priors
described in
Section~\ref{sec:priors}. Specifically, we considered values of 1 and 2
for the upper bound $u$ in the uniform prior for the variability
parameters $\lambda$s,
considering that the overall standard deviation in the data is 0.5. We
tried all the
three options for generating the mean vectors. We computed summary
statistics from the data to use as the parameters in the OU process and
Brownian motion. For example, we used the sample mean and standard
deviation of the data at 0 min as the mean and standard deviation,
respectively, of the normal distribution we assume for the starting
values of the OU process or the Brownian motion. We also compared the
Gibbs and the MH samplers for the concentration parameter $\alpha$.
These different choices turned out not to have much impact on the results.

To examine the sensitivity of our method to changes in the data, we
applied DIRECT also to the larger data set of 270 genes, identified at
an FDR of 25\% by EDGE.
DIRECT identified 19 clusters for this larger data set [Figures~1--3 in
the supplemental material; \citet{Fu.etal.2013.supp}].
The cluster allocation is similar to that for the 163 genes, with the
additional 107 genes allocated to most of the clusters identified for
the 163 genes
[Figures~1--3 in the supplemental material; \citet{Fu.etal.2013.supp}].

\subsection{Biological implications}
The inferred clustering suggests roughly\break three stages of gene
expression in response to a pulse of Notch activation (Figure~\ref{fig:de163.curves}): before 50 min (early response),
between 50 and 100 min (late response), and around and after 100 min
(very late response). Clusters 9 and 12 showing early transcriptional
responses contain
most of the known target genes, that is, Notch has a direct impact on
the transcriptional changes of these genes.
Cluster 7 showing late responses also contains 3--5 known targets~[\citet
{Krejci.etal.2009}], but approximately 10 other genes in this cluster
may also be
Notch targets. Genes in other late or very late response clusters may
be Notch targets as well. Together with our collaborators, we analysed
data from
additional experiments to examine whether this is the case~[\citet
{Housden.etal.2012}]. Furthermore, it is known that Notch generally
promotes transcription rather than represses
it, and that the early-upregulated genes in cluster 9 are strong
repressors. Our clustering therefore suggests unknown,
complex regulation mechanisms involving interactions between different
clusters of genes. With additional experiments, \citet
{Housden.etal.2012} investigated possible transcriptional regulation
mechanisms and identified a feed-forward regulation relationship among
clusters 9, 6 and 7.

\subsection{Results from SplineCluster and MCLUST}
For comparison, we ran SplineCluster and MCLUST on the two real data
sets, using the average profiles and the default settings (Table~\ref{table:notch.nclust}).
SplineCluster inferred only 7 clusters for both data sets, with the
inferred clusters exhibiting a much higher level of heterogeneity than
under our DIRECT method
[Figures~4--5 in the supplemental material; \citet{Fu.etal.2013.supp}].
This result is consistent with its performance on simulated data:
SplineCluster also tends to infer
a lower number of clusters in case of high heterogeneity (Section~\ref{sec:simulations} and Table~\ref{table:simu.compare.2}). MCLUST
inferred 15 clusters for
163 genes, which is comparable to our DIRECT method [Figures~6 and 8 in
the supplemental material; \citet{Fu.etal.2013.supp}]. However, it
inferred only 2 clusters
for 270 genes and a different covariance model [Figures~7 and 9 in the
supplemental material; \citet{Fu.etal.2013.supp}].
This sensitivity of clustering to the relatively minor change in the
data may have arisen from MCLUST trying to simultaneously select the
number of clusters
and the covariance model. Selection of the covariance model adds
another layer of complexity to the problem of clustering, particularly
when none of the
different covariance models considered by MCLUST is compatible with the
experimental design. The uncertainty in the covariance model selection
may also
explain the particularly high variability in the inferred number of
clusters for simulated data in simulation \#1 (Table~\ref{table:simu.compare.2}).
%
\begin{table}
\tablewidth=250pt
\caption{Numbers of clusters estimated by three clustering methods:
DIRECT, SplineCluster and MCLUST for genes identified by EDGE~[Storey et~al.
(\citeyear{Storey.etal.2005})]
to be differentially expressed over the time course}
\label{table:notch.nclust}
\begin{tabular*}{250pt}{@{\extracolsep{\fill}}lcc@{}}
\hline
& \multicolumn{2}{c@{}}{\textbf{No. of inferred clusters}} \\[-6pt]
& \multicolumn{2}{c@{}}{\hrulefill} \\
& \textbf{163 Genes} &\textbf{ 270 Genes} \\
& \textbf{(FDR 10$\bolds{\%}$)} & \textbf{(FDR 25$\bolds{\%}$)} \\
\hline
DIRECT & 14 & 19 \\
SplineCluster & \phantom{0}7 & \phantom{0}7 \\
MCLUST & 15 & \phantom{0}2 \\
\hline
\end{tabular*}
\end{table}

\section{Discussion}
\label{sec:discussion}
In this paper we developed DIRECT, a model-based\break Bayesian clustering
method for noisy, short and replicated time-course data. We implemented this
method in the R package DIRECT, which may be downloaded from CRAN (\url
{http://cran.r-project.org/web/packages/}). We also applied this method
to analyse the time-course microarray gene expression levels following
Notch activation in \textit{Drosophlia} adult muscle cells. Our analysis identified
14 clusters in 163 differentially expressed genes and assigned
probabilities of cluster membership for each gene. The clustering
results indicate
three time periods during which genes attain peak up- or
down-regulation, which was previously unknown, and suggest
possibilities for the
underlying mechanisms of transcription regulation that may involve
interactions\vadjust{\goodbreak} between genes in different clusters. Hypotheses on the
biological mechanisms are further
investigated in \citet{Housden.etal.2012}. Here we discuss several
additional aspects of the clustering method.

Our method has four main features.
First, the random-effects mixture model decomposes the total
variability in the data into three types of variability that arise from
clustering (${\lambda_\phi}$), from sampling
across multiple experimental conditions (${\lambda_\tau}$), and from sampling a
limited number of replicates (${\lambda_\varepsilon}$). This variance decomposition
regularises the
covariance matrix with constraints that are consistent with the
experimental design. It is simultaneously parsimonious and identifiable
for the replicated data:
the replicated profiles at multiple time points of a single gene are
already informative for ${\lambda_\tau}$ and ${\lambda_\varepsilon}$, and having at least 2 genes in
a cluster makes ${\lambda_\phi}$
estimable. Second, our
method uses the Dirichlet-process prior to induce a prior distribution
on clustering as well as the number of clusters, making it possible to
estimate directly
both unknowns from the data. Third, we have developed a novel
Metropolis--Hastings MCMC algorithm for sampling under the
Dirichlet-process prior.
Our MH algorithm allows the use of nonconjugate priors. It is also
efficient and accurate, as simulation studies demonstrate. Fourth, our
method infers the
posterior allocation probability matrix through resampling and
relabeling of the MCMC samples. This probability matrix can then be
used directly in forming
clusters and making probabilistic cluster allocations. Simulation
studies and application to real data show that DIRECT is sensitive
enough to variability in
the data to identify homogeneous clusters, but not too sensitive to
minor changes in the data.

Several other model-based clustering methods construct their models
along similar lines~[\citet{Celeux.etal.2005, Ma.etal.2006,
ZhouWakefield2006, Booth.etal.2008}].
In fact, our model in equation~(\ref{eqn:rem}) coincides with the
random-effects model E3 in \citet{Celeux.etal.2005}. However, those
authors decided to focus on a
slightly simpler model, which is similar to equation~(\ref{eqn:rem})
but without the within-component random effects $\phi^k_i$. They based
their decision on the
nearly identical likelihoods of the two models for simulated data. \citet
{Ma.etal.2006} and \citet{ZhouWakefield2006} did not deal with
replicated data and included in their model
only two types of variability: the within-cluster variability and the
variability due to multiple time points.
Similar to us, \citet{Booth.etal.2008} worked with replicated
time-course data and used random effects to account for different types
of noise, but
their partition of the total variability is not based on the
experimental design and is therefore much less straightforward.
Specifically, they allowed for dependence
among different items in the same cluster but did not explicitly
account for the random effect due to time (or experimental condition).

Note that our DIRECT method does not account for the temporal
structure, but rather focuses on modeling the covariance matrix.
This approach is similar to
MCLUST, which applies\vadjust{\goodbreak} eigenvalue decomposition to the covariance matrix
and considers various constraints on the decomposed
covariance matrix (i.e., whether the shape, orientation or volume of
the covariance matrix is identical across mixture components), although
the constraints
considered in MCLUST are not based on any experimental design. The good
performance of
our method on both simulated and real data, and of MCLUST in several
cases, suggests that accounting for the temporal structure in the mean
vectors, such
as via splines functions as in SplineCluster or via Gaussian processes
as in \citet{ZhouWakefield2006} and others, may not be necessary. We
also followed the
approach in \citet{ZhouWakefield2006} and modeled the mean vector of
each mixture component as a Brownian motion (with drift) and, extending
this idea,
as an Ornstein--Uhlenbeck process. The clustering results such as the
inferred number of clusters and allocation of individual genes did not
change much,
because these approaches impose the temporal structure on the mean
vector: conditioning on the correct clustering, the data are directly
informative
of the cluster mean, a main parameter of interest.
Incidentally, DIRECT is applicable also in more general cases of
multiple experimental conditions, where dependence among conditions is
nonexistent, unclear or unknown.

Similar to other MCMC methods, our DIRECT method does not aim to
optimise the runtime. Whereas MCLUST and SplineCluster, both non-MCMC methods,
took only seconds or at most minutes to run on the simulated and real
data here, we ran DIRECT for hours to ensure the consistency in results across
different runs, which indicated that the Markov chain had mixed well.

We have used only the one-parameter Dirichlet-process prior in our
method. The concentration parameter in the Dirichlet-process prior
simultaneously controls the
number of clusters as well as the size of each individual cluster. The
prior has the tendency of creating clusters of very different sizes.
The posterior inference
to generate the posterior allocation probability matrix is therefore
critical to balance out the unevenness: although certain clusters may
be very small or very
big in a single iteration, items allocated to these tiny clusters are
likely allocated to other, possibly larger, clusters over a sufficient
number of MCMC iterations. Nonetheless,
as pointed out by the Associate Editor and an anonymous reviewer, other
exchangeable priors, such as the two-parameter Dirichlet process [aka
the Pitman-Yor process; \citet{PitmanYor1997}] and many
other extensions of the Dirichlet process reviewed in \citet
{Hjort.etal.2010}, may also be adopted under our framework. Indeed,
these other exchangeable priors
may offer more flexibility and suggest an important direction to extend
our current work.

Under our and \citet{Neal2000}'s MH MCMC algorithms, the Markov chain is
constructed for the cluster memberships of individual items.
Generation of a new cluster and elimination of an existing cluster are
implied rather than enforced.
In contrast, reversible-jump MCMC~[\citet{RichardsonGreen1997}] and\vadjust{\goodbreak}
birth-death MCMC~[\citet{Stephens2000a}] enforce changes in dimensions
by designing the MCMC moves around the number of clusters. Their
strategy may not be efficient for clustering multivariate data, because
even a fixed number of clusters may correspond to a large number of
possible partitions and a large space
of the cluster-specific parameter values. For clustering it seems more
sensible for the Markov chain to move to the neighbourhood of the
``correct'' number of
clusters and to fully explore the parameter space in this
neighbourhood, as under Neal's approaches and under our method.

\begin{appendix}\label{sec:app}
\section*{\texorpdfstring{Appendix: Proof of Proposition \lowercase{\protect\ref{prop:condprob}}}{Appendix: Proof of Proposition 1}}

We use the joint distribution of clustering and the number of clusters
given in equation (\ref{eqn:dir.prior}) for derivation. Let $K_{-i}$ be the
number of clusters when the $i$th gene is excluded. Then,
\begin{eqnarray*}
&&\Pr(Z_i=z, K=k| \Z_{-i} = \z_{-i}, \alpha) \\
&&\qquad=
\frac{\Pr(\Z=\z, K=k |
\alpha)}{\Pr(\Z_{-i}=\z_{-i}, K_{-i}=k_{-i} | \alpha)}
\\
&&\qquad= \frac{\Gamma(\alpha) / \Gamma(\alpha+N) \alpha^k \prod^k_{l=1}
(N_l-1)!}{\Gamma(\alpha) / \Gamma(\alpha+N-1) \alpha^{k_{-i}} \prod^{k_{-i}}_{s=1} (N_s-1)!}
\\
&&\qquad= \cases{ \displaystyle\frac{N_{z} -1}{N-1+\alpha}, &\quad $\mbox{$Z_i$ is not in a singleton
cluster}, $\vspace*{2pt}
\cr
\displaystyle\frac{\alpha}{N-1+\alpha}, &\quad $\mbox{$Z_i$ is in
a singleton cluster}.$}
\end{eqnarray*}
Alternatively, \citet{Neal2000} derived the above result first under the
finite mixture model, treating $K$ as a constant, and then letting
$K\rightarrow\infty$.
\end{appendix}

\section*{Acknowledgements}
The authors thank Ben Housden, Alena Krejci and Bettina Fischer for
collecting and sharing the time-course microarray gene expression
data analysed here. AQF also thanks Jurgen Van Gael, Richard Samworth,
Richard Nickl, Pat Altham and Matthew Stephens for helpful discussions
on statistics and to Robert Stojnic and Laurent Gatto for advice on R
programming. Thanks also go to Editor Dr. Karen Kafadar, the Associated
Editor and
two anonymous reviewers, whose thorough and insightful comments greatly
improved the manuscript.


\begin{supplement}[id=suppA]
\stitle{Appendices}
\slink[doi]{10.1214/13-AOAS650SUPP} 
\sdatatype{.pdf}
\sfilename{aoas650\_supp.pdf}
\sdescription{The pdf file contains the proof of Proposition \ref{prop:hastings}, details
on the MCMC algorithm and additional figures.}
\end{supplement}

%

\printaddresses

\end{document}